\documentclass[twoside,english,12pt]{Cern2}

\usepackage{amssymb}
\usepackage{graphicx}
\usepackage{babel}
\usepackage{multicol}
\usepackage[centerlast]{caption2}

\begin{document}

\title{Limitations on the Use of Acoustic Sensors in RF Breakdown Localization}

\author{F. Le Pimpec\thanks{E-mail: lepimpec@slac.stanford.edu} \\
SLAC, 2575 Sand Hill Road Menlo Park CA 94025 , USA}

\maketitle

\thispagestyle{headings} \markright{SLAC-TN-04-049 $\backslash$
LCC-0149}

\begin{abstract}

X-band accelerator structures, meeting the Next Linear Collider
(NLC) design requirements, have been found to suffer damage due to
radio frequency (RF) breakdown when processed to high gradients
\cite{lep_otpsy:Linac02}. Improved understanding of these
breakdown events is desirable for the development of structure
designs, fabrication procedures, and processing techniques that
minimize structure damage. Using an array of acoustic sensors, we
have been able to pinpoint the location of individual breakdown
events. However, a more accurate localization is required to
understand the interaction between the phonon or the sound wave
with the OFE copper.

\end{abstract}



\section{Introduction}
As part of the R\&D effort for the Next Linear Collider (NLC), the
attainment of high gradients (70MV/m) with a breakdown rate below
1 per 10 hours must be demonstrated \cite{snowmassLC} \cite{zdr}.
In the Next Linear Collider Test Accelerator (NLCTA), at SLAC
(Stanford Linear Accelerator Center), RF travelling and standing
wave copper structures, designed to meet the needs of the NLC
\cite{adolphsen:pac97}, are being tested.

\begin{figure}[tbph]
\begin{center}
\includegraphics[width=0.8\textwidth,clip=]{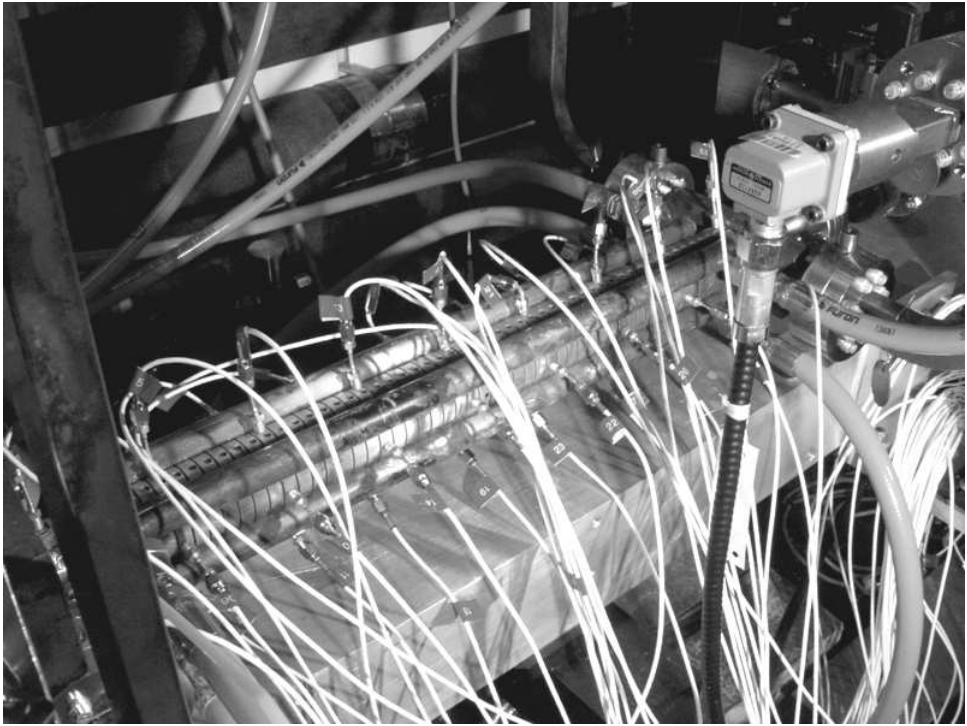}
\end{center}
\caption{T53VG3RA (53cm long travelling wave structure of 3\%
group velocity) structure covered with acoustic sensor}
\label{AET53RAoverall}
\end{figure}

\medskip
To reach an NLC accelerating field of 70MV/m with a 400ns pulse
length, megawatts of RF power are poured into the structures.
Depending on the design and type of structure, this power can vary
from 73MW for a 60cm long travelling wave structure with a 3\%
group velocity (H60Vg3) to 150MW for some structures 180cm long.
Part of the RF power is transformed in the copper into heat. The
lost power is up to 2/3 of the input power for a structure kept at
45$^\circ$C. The thermal expansion of the copper, as the
structures fill with RF power, causes sound to occur on every RF
pulse. Using extremely sensitive piezoelectric microphones
(acoustic sensors), it is possible to "listen" to the accelerator
structure as it is running, cf Fig.\ref{AET53RAoverall}.

\medskip
High gradients are obtained by exposing the structure to High
Power Pulsed RF (HPP). This technique is called RF processing.
During processing, the structure occasionally responds by arcing,
i.e breakdown. A breakdown is characterized by a shut off of the
transmitted power with up to 80\% of the incident power absorbed
in the arc \cite{dolgashev:linac2002}. This extra energy is
deposited in the copper, and a part of it is converted in extra
phonons (heat and acoustic) that can be picked up by our acoustic
sensors. With this technique, crude localization of a breakdown is
straightforward, and complementary to the standard RF analysis
with directional RF couplers.

\medskip
Getting a precise localization of where the breakdown occurs
requires a short wavelength sound wave. However, the interaction
between the wave and the copper grain structure can become
troublesome, hence not allowing good spatial resolution. In order
to go forward on localizing breakdowns, we tried to identify the
problem cause(s).

\section{From RF Breakdown to Acoustic Sensor}
\label{RFbrkACsensor}

The RF accelerating structure is subject to chemical degreasing
and etching and thermal treatment, this step is the
pre-processing, before it is RF tested. The use of acoustic
sensors in the environment of accelerators is rather new and the
way of pre-processing a structure has an impact, not only on the
performance of an RF structure, but also on the understanding of
the acoustic results obtained, cf \S\ref{Acousticphysic}. Despite
the fact that detailing the cleaning procedure is out of the scope
of this paper, it is still necessary to mention the SLAC process,
as no suitable reference exists.

\medskip
The NLC cleaning procedure for most of the structures built so far
starts with a perchlorethylene vapor degreasing for all single OFE
copper cells, followed by a chemical etching (up to 60s for
poly-crystalline machined cells). The cells are assembled into
structures and bonded at 1020$^\circ$C in an atmosphere of
hydrogen, the full structure is "wet" and "dry" hydrogen fired at
950$^\circ$C. Ancillary couplers and water cooling tubes are then
brazed onto the structure. The structure is then RF tuned using a
bead pull technique \cite{Hanna:pac97}. Finally the hydrogen is
thermally desorbed by vacuum firing for two weeks at 650$^\circ$C.
The full structure is then installed at the NLCTA and baked at
220$^\circ$C, before the beginning of an RF processing. The
purpose of this extreme fabrication/processing schedule is the
removal of the contaminants, gas and particles, that contribute to
field emission and participate in breakdown. The in-situ bakeout,
at 220$^\circ$C, is now omitted. FNAL (Fermi National Laboratory)
has a similar process but the structure are brazed in a 500~mTorr
Ar atmosphere, and then vacuum fired \cite{Arkan:Epac04}.

\medskip
There is two ways to RF process a structure, one is to slowly
apply an electric field with a high enough intensity to
electrostatically remove a particle from a surface. The length of
the pulse has to be short enough for avoiding the fusion, by Joule
heating, of the particle to the surface. The second way, used for
NLC structures, is to apply a very intense electric field in order
to melt the emitter, without causing damage to the accelerating
structure \cite{tan} \cite{llaurent}. The RF processing starts
with fixed pulses of 50~ns, and pedestal electric field. The
electric field inside the structure is raised every minute by 1 to
2\% until it reaches 70 to 75~MV/m, unless a breakdown happen.
This sequence is then repeated for pulses length of 100~ns,
170~ns, 240~ns and 400~ns. If a breakdown occurs during the pulse,
the RF control system shuts off for 60~s. When resuming operation
the RF control system is ramps up, first with the power at a
shorter pulse length and then is widening the pulse. Depending on
the structure, this RF processing can take from 100 to 1000 hours.

\medskip
Despite such careful pre-processing, breakdown can occur during RF
processing. These breakdowns result in physical damage to the
structure \cite{lep_otpsy:Linac02} and more troublesome, detuning,
or phase shift \cite{adolphsen:pac01}. The particle beam, electron
or positron is fully accelerated at every cell of the RF structure
only if the beam is at the right phase of the accelerating
electrical field. The structure is designed in such a way that
each cell has a geometry that insures resonance at 11.424~GHz and
appropriate phase advance. The detuning of a single cell due to
physical damage is a change in the geometry of the cell. If the
phase shift is too large, the acceleration is not only less
efficient but the emittance (size of the particle beam) cam also
be affected. However, physical damages in the structure are
irrelevant as long as the overall structure accelerates properly.
In order to avoid damages, and detuning, it is vital to understand
the breakdown mechanism.
\newline DC breakdown mechanism studies \cite{latham:1995} have started
approximatively 50 years, long before RF study, and so far no
strong correlation have been found between the breakdown mechanism
in RF and DC \cite{llaurent}. Hence, we need to add new techniques
to old ones to enhance our understanding of RF breakdown.

\medskip
The general hypothesis used to explain breakdown, RF or DC, is
based on field emission. The intense electric field present at the
surface of the irises of the cells draws an electron current. This
current will locally ionize residual gas as well as heat the
surface, which release more gas, and producing more ions and
electrons. These processes happen at every RF pulse, without
causing a systematic arc. The formation mechanism of an arc, due
to a build up of electron and ion current, for a single pulse is
not yet clear. However, the characteristics of a breakdown are
rather well understood \cite{dolgashev:linac2002}. In some cases,
such as at the input coupler of some of the NLC structures, the
breakdown is not initiated by the electric field but by the
induced current along sharp edges due to the high magnetic field
\cite{dolgashev:linac2002} \cite{pritzkau}. The route toward
developing a stable NLC structure pass either by a change of cell
material (Mo, W, eventually stainless steel), such as the
structure used at CERN (Conseil Europ\'een pour la Rcherche
Nucl\'eaire) \cite{dobert:linac2002}, or/and by a better
understanding of breakdown itself \cite{llaurent}
\cite{Norem:2003} \cite{Wilson:pac03}.

\medskip
One of the properties of an RF breakdown, as we define it for our
application, is characterized by the fact that 80\% of the
incident power is absorbed by the arc \cite{dolgashev:linac2002}.
Simulations, based on plasma spot model, performed to better
understand the mechanism of breakdown, do not account for all the
loss of energy. A part of this energy is converted into an ion
current. Light from excited neutral copper (Cu I) has been
observed with a spectrum analyzer at SLAC but also at CERN on
similar experiments \cite{dobert:linac2002}. The reflected RF also
accelerates some electrons toward the upstream cells, this current
is known as the dark current. Finally, a part of the energy is
converted to heat inside the structure, inducing stress which
produces sound waves.\newline Several experiments have been
conducted in the NLCTA to account for the missing energy, during a
breakdown, with mixed results. Those experiments included
radiation monitoring, dark current and acoustic measurements.

\medskip
In order to locate breakdowns in superconducting RF accelerating
cavities, thermal sensors are used \cite{knobloch}. Heat deposited
during arcing is recorded by thermometers. Heat is carried by high
frequencies phonons (1~GHz to 1~THz), also called optical phonons.
Using the same idea for warm (room temperature) accelerating
structure, we detect sound wave or low frequency acoustic phonons
(100 KHz up to 20 MHz range).

\medskip
At room temperature (RT), high frequency phonons have a mean free
path smaller than their wavelength. As a result heat is propagated
in a diffusive way and is described by the heat diffusion equation
(equ.\ref{heattransfer}).

\begin{equation}
\rho \, C_p \, \frac{\partial T}{\partial t} = Q + \nabla.(k
\nabla T) \label{heattransfer}
\end{equation}

where $\rho$ is the density,  C$_p$ is the heat capacity at
constant pressure, $\frac{\partial T}{\partial t}$ is the change
in temperature over time, Q is the heat added, k is the thermal
conductivity.

\medskip
However, at low temperature (2K), high frequency phonons can
propagate in a ballistic way (straight line). This property might
be one of the reasons why thermometry gives not only a precise
localization for the RF breakdown, but also can locate the active
field emitter on the surface of the RF cavity \cite{knobloch},
before breakdown. At room temperature, low frequency acoustic
phonons also propagate in a ballistic way. Their propagation
follows the wave equation (equ.\ref{waveequat}).

\begin{equation}
\nabla ^2\psi = \frac{1}{v^2} \ \frac{\partial \psi ^2}{\partial
t^2} \label{waveequat}
\end{equation}

where $v$ is the speed of propagation of the wave and t the time.

\medskip
Very good results are obtained for superconductivity cavities with
the thermometry method, therefore it is reasonable to try an
equivalent technique. We utilized, for our experiment, the same
kind of ultrasonic sensor which is used in non-destructive testing
\cite{AEStesting:87} \cite{krautk:90}.

\medskip
In some cases, several breakdowns can occur on a single machine
pulse. We name these events double breakdown, and so on. Such
events are difficult to locate with conventional RF techniques,
using directional couplers. We will show that the use of an array
of acoustic sensors provides an easy way to account for a more
accurate number of breakdown as well as their localization. On the
end, this is also a means toward understanding the underlying
causes of these arcs.

\section{Copper and Acoustic physic}
\label{Acousticphysic}

\begin{figure}[tbph]
\begin{center}
\includegraphics[width=0.8\textwidth,clip=]{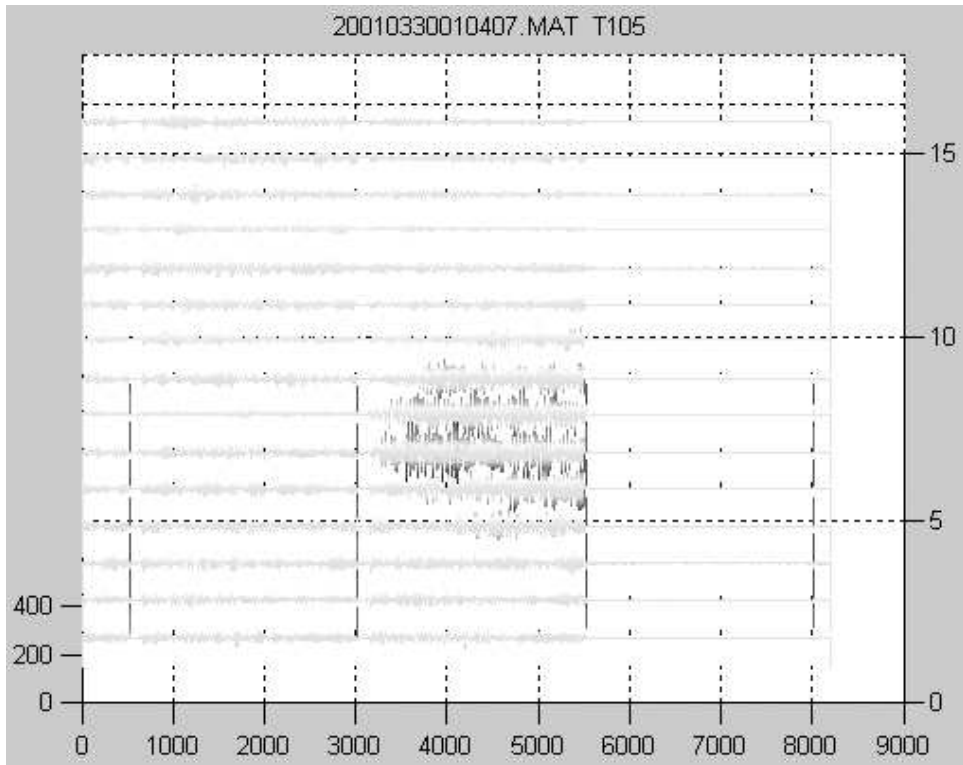}
\end{center}
\caption{Acoustic bow wave in the T105VG5 accelerating structure following an RF
breakdown. Sensors are located 6 cells ($\sim$5cm) apart on the
spine of the structure, cf Fig.\ref{AET53RAoverall}}
\label{T10510407}
\end{figure}

The propagation of a sound wave is described by the wave equation
(equ.\ref{waveequat} $\S$\ref{RFbrkACsensor}). The phonons travel
at the speed of sound in the medium and, if the medium is
isotropic, the wave can be easily constructed using the classical
\emph{Huyghens principle}: "Every point on a propagating wavefront
serves as the source of spherical secondary wavelets, such that
the wavefront at some later time is the envelope of these
wavelets. For a propagating wave of a frequency, f, transmitted
through a medium at a speed, v, the secondary wavelets will have
the same frequency and speed". Following this principle, the wave
takes a spherical shape. If a few sensors are placed on the same
plane, the sensors will respond one after the other and a plot of
the response versus time of the sensors will show a bow wave.
\newline An example of such a response from sensors after a single RF
breakdown in a travelling wave structure of 105 cm long, 5\% group
velocity (VG), is displayed in Fig.\ref{T10510407}. The abscissa
is equivalent to time in arbitrary units. The time difference
between 2 points is of 0.1 $\mu$s. The left ordinate is a level of
the amplitude also in arbitrary units. The right ordinate is the
sensor number. The plot is separated in 3 regions representing 2
consecutive RF pulses and 1 after pulse. The RF breakdown happens
in the second pulse and, as a result, the RF power is shut off and
no energy is present on the third pulse.

\medskip
However, if the medium is anisotropic, the wave vector and the
group velocity vector are no more collinear. The construction of
the envelope of the wave becomes more complicated. The anisotropic
propagation of elastic energy in a material having different
crystal orientation is known as phonon focusing \cite{Wolfe:98}
\cite{Wolfe:92}.\newline The primary idea behind phonon focusing
is that phonons are wavepackets that travel at the group velocity.
In any anisotropic media, an isotropic (in k-space) phonon
distribution will result in an anisotropic distribution of phonon
flux because of anisotropies in the group velocity.  This is seen
as a concentration of thermal energy along particular crystal
directions. This concentration is easiest to observe in pure,
single crystals because inelastic scattering events can randomize
the direction of travel of a wavepacket, making the propagation
appear isotropic.  However, since the phonon scattering decreases
with decreasing frequency, it is possible to observe focusing of
low frequency phonons, even in imperfect crystals \cite{msall}. As
a result two sensors located near each other might not receive the
same amounts of energy \cite{Wolfe:98}. An example of the
existence of the phonon focusing effect in single copper crystal
with a 15MHz transducer is shown in \cite{Wolfe:98}.

\medskip
Acoustic phonons are, in the first approximation, propagating
ballistically, compared to optical phonon which propagate
diffusively. The speed of sound in copper, associated to the
propagation of the longitudinal wave also called compression wave
or pressure wave (P-wave), is given by :

\begin{minipage}[h]{.5\linewidth}
\centering
$$
 v = \sqrt{\frac{E}{\rho}}=3560 m/s
$$
\end{minipage}
\begin{minipage}[h]{.5\linewidth}
\centering
\begin{equation}
\left\{ \begin{array}{ll}
    \mbox{E : Young modulus (Pa)} \\
    \rho = 8930 \ kg/m^3
    \end{array}
    \right.
 \end{equation}
\end{minipage}

\medskip
Inside or at the surface of the copper structure, other sound
waves might also propagate. These other waves are shear waves or
transverse waves (S-wave or T-wave), and surface waves
(Rayleigh-Lamb Wave). In a longitudinal wave, the displacement of
the medium (or atoms) is parallel to the propagation of the waves.
For shear wave the displacement of the medium is perpendicular to
the propagation of the wave. Surface waves propagate at the
interface between two media as opposed to through a medium. Many
types of surface waves can exist and for our purpose we will
consider only the Rayleigh-Lamb waves (R-waves). Rayleigh waves have
longitudinal and shear displacement coupled together which travel
at the same velocity. The medium oscillates in an elliptical path
within the vertical plane containing the direction of the wave
propagation \cite{encyclopedia} \cite{Schlumberger}
\cite{David:2002} \cite{Weisstein}.
\newline The copper used for building our RF structures is
annealed. From the data of the speed of sound in annealed copper
\cite{crc}, we can determine the Young modulus, the poisson ratio
(usually $\nu \sim$0.3 for copper). Plugging those values in the
following equations, we can calculate the Transverse $ v_S$ and
the Rayleigh wave $v_R$ velocities. Equation (\ref{rayleighwv}),
from Bergmann, gives an approximate value for the Rayleigh wave
velocity \cite{AEStesting:87} \cite{krautk:90} \cite{Bachm:96}

\begin{minipage}[h]{.5\linewidth}
\
\begin{equation} v_p = \sqrt{\frac{E \ (1 - \nu)}{\rho \ (1 + \nu)(1 - 2\nu)}}
\label{pressurewv} \end{equation}
\begin{equation} v_s = \sqrt{\frac{E}{2\rho \ (1 + \nu)}} \label{shearwv} \end{equation}
\end{minipage}
\begin{minipage}[h]{.5\linewidth}
\begin{equation} v_R \simeq \frac{v_s \ (0.86 + 1.14\nu)}{1 + \nu} \label{rayleighwv} \end{equation}
\end{minipage}

The speed of the longitudinal wave $v_p$ and the shear wave $v_s$
are respectively 4760~m/s and 2325~m/s \cite{crc}. The
determination from equations (\ref{pressurewv}) and
(\ref{shearwv}) gives a Rayleigh wave speed of $\sim$2160~m/s. The
equations written above can also be expressed as a function of the
Lam\'e coefficients $\lambda$ and $\mu$

\begin{minipage}[h]{.5\linewidth}
\begin{equation} \lambda = \sqrt{\frac{\nu E}{(1 + \nu)(1 - 2\nu)}}
\label{lamelambda} \end{equation}
\end{minipage}
\begin{minipage}[h]{.5\linewidth}
\begin{equation} \mu = \sqrt{\frac{E}{2 \ (1 + \nu)}} \label{lamemu} \end{equation}
\end{minipage}

\medskip
The difference in speed of these waves might in principle be used
to determine the location of a breakdown. In the same way
seismologists localize earthquakes. The distances between the
breakdown and the sensors is of the order of a few cm. The time
between the front edge of the pressure wave and a shear wave is in
the order of a few microseconds. Electronics are fast enough to
separate the edges in time. The problem lies in the energy carried
by the waves and the ability to separate the arrival of a
successive wave from the tail of the preceding wave. Fast arrival
of successive waves might mimic a single wave with long ringing,
as in Fig.\ref{T10510407}.

\medskip
Depending on the spatial resolution and the precision needed to
localize the source of an RF breakdown, going to higher
frequencies might be necessary. As the frequency goes up,
$\sim$10-20~MHz, the wavelength of the acoustic wave might reduce
to the size of a grain (0.5~mm to 0.3~mm for the pressure waves).
Grain size in our RF structures (annealed copper) can be up to a
millimeter.
\newline Thermal processing (cf $\S$\ref{RFbrkACsensor}) and exposure
to high pressure of hydrogen changes the way the grains grow and
further, the intake of hydrogen might create voids, by reaction
with oxide inclusions, at the grain boundaries and dislocations
\cite{butt:83} \cite{Nakahara:88} \cite{brongersma:2002}. This
intake of hydrogen inside the bulk of the material might
ultimately affect the propagation of ultrasonic waves
\cite{Zielinski:90}, if not properly degassed. However, even if
the copper is well degassed, the temperature of the vacuum firing
might not be high enough to remove the dislocation or voids
created by the intake of hydrogen. Meeting with experts did not
rule out this hypothesis \cite{h2material:02}. An understanding of
the effect of dislocations and grain boundaries on the propagation
of ultrasound waves is necessary, as is the attenuation of those
waves inside a medium.

\medskip
Geometric attenuation or attenuation due to the spread of the wave
is different for a bulk wave or a surface wave. The bulk waves
attenuate following a $1/r$ law where surface wave attenuates as
$1/\sqrt(r)$, for a detector located far from the source
\cite{AEStesting:87}. In addition to the geometric attenuation,
the wave can scatter and diffract \cite{AEStesting:87}
\cite{krautk:90}. Scattering of the sound wave will occur for
inhomogeneous material. Grain boundaries and defects like
inclusions and voids are interfaces where an abrupt change of
impedance for acoustic waves occurs. This abrupt change create a
reflected and a transmitted wave from the  initial incoming wave.
As a rule of thumb, scattering of sound waves having a wavelength
a hundred to a thousand times larger than the grain size is
negligible. The scattering is important for waves having a
wavelength of the dimension to 10 times the dimension of the grain
\cite{krautk:90}. Finally, energy attenuation can be due to
natural absorption by the medium of the acoustic energy. It can
happen by conversion of kinetic energy into heat, by loss of
energy through plastic deformation, interactions with dislocation
motion. Losses can also occurs by friction between surfaces that
slip \cite{AEStesting:87}. We have observed during autopsy of some
RF structures, cf. \cite{lep_otpsy:Linac02} and \cite{Harvey:AVS},
slip bands. This matter was investigated separately. Also in order
to be complete, electron, ion, or phonon - phonon interaction can
be sources of loss of energy. These sources of losses tend to be
important for phonons of much higher frequencies than the ones
with which we are working.

\begin{figure}[tbph]
\begin{center}
\includegraphics[width=0.7\textwidth,clip=]{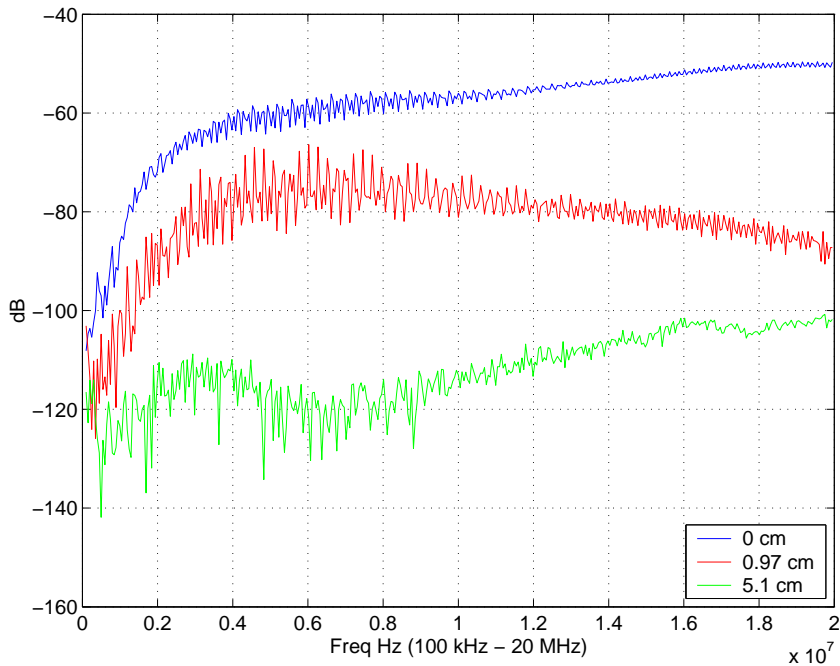}
\end{center}
\caption{Attenuation of the acoustic response in a Copper block of
0-5~cm thickness with a 20MHz sensor driving \& 15MHz sensor
receiving}
\label{Cuatenuation}
\end{figure}

\medskip
In order to understand how different acoustic waves propagate in
annealed copper, a collaboration with the University of Illinois
\cite{Gollin} on an experimental and modelling program have been
started. This collaboration will answer questions started by few
experiments done at SLAC \cite{Greenwood:2002}. In these
experiments, using an un-annealed copper block of approximatively
10~cm x 10~cm x 5~cm, we found out that high frequencies (above
1-2~MHz) did not propagate well through the block. In the setup, a
spark or a laser was shot in one of the side of the block and the
sound signal was recorded on the opposite side by an acoustic
sensor. Fig.\ref{Cuatenuation} shows the response of an acoustic
sensor receiving an ultrasonic signal peaked at 20MHz travelling
through a copper plates of different thicknesses.

\begin{figure}[tbp]
\begin{minipage}[t]{.5\linewidth}
\centering
\includegraphics[width=0.9\textwidth,clip=]{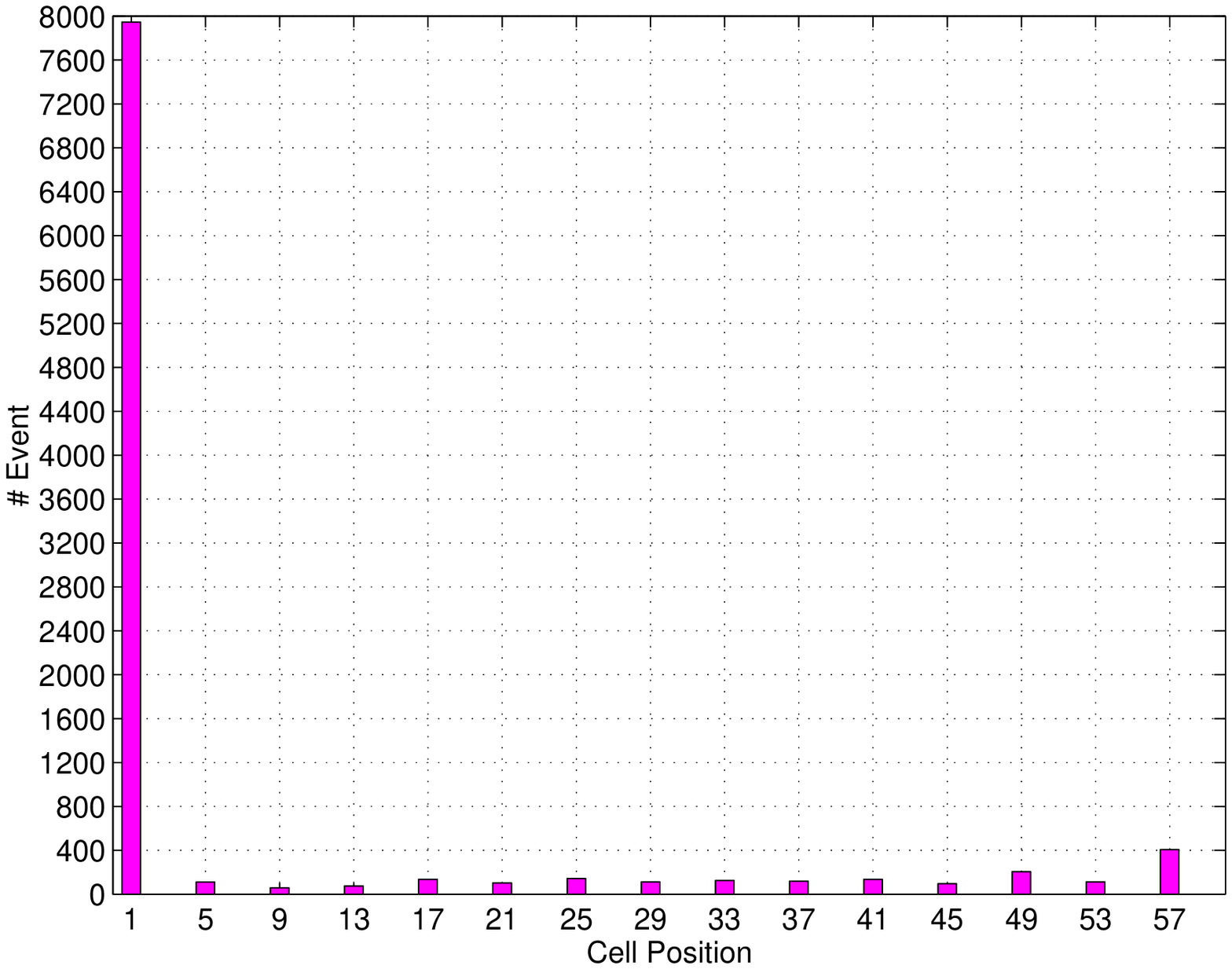}
\end{minipage}%
\begin{minipage}[t]{.5\linewidth}
\centering
\includegraphics[width=0.9\textwidth,clip=]{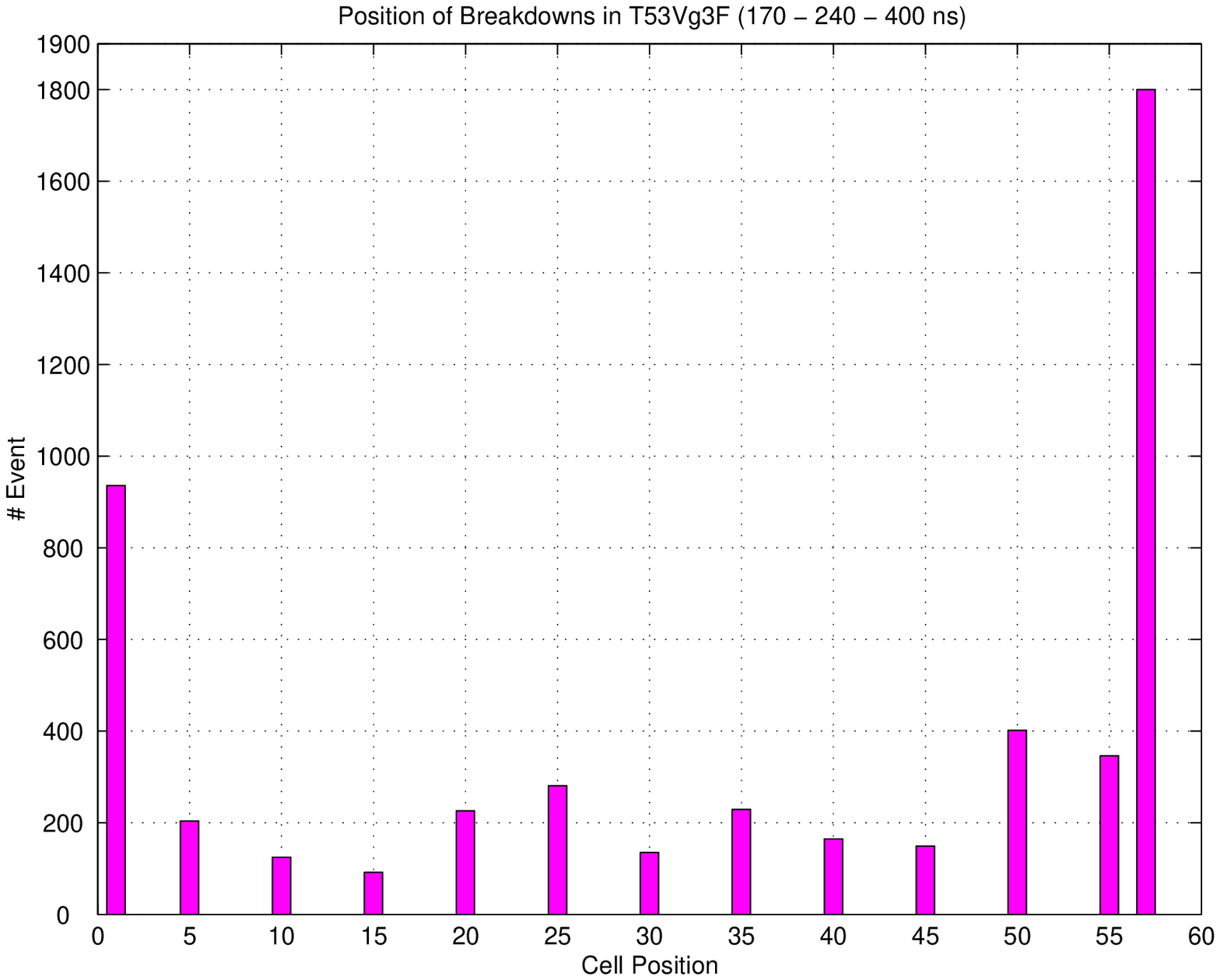}
\setcaptionwidth{4cm}
\end{minipage}
\caption{Number of single breakdown vs location of the 53 cm long
RA (left) and F (right) travelling wave structure; for RF pulses
of 170ns 240 ns and 400ns length}
\label{T53RAFbrkdwn}
\end{figure}

\begin{figure}[tbph]
\begin{center}
\includegraphics[width=0.8\textwidth,clip=]{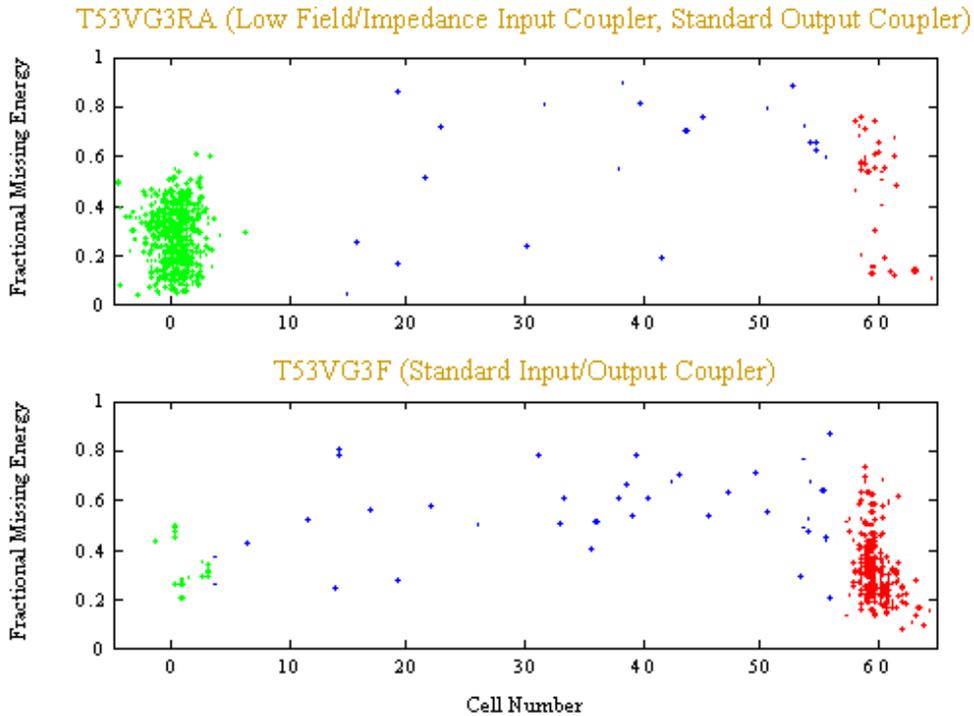}
\end{center}
\caption{Localization of RF breakdown by the use of RF analysis.
120 Hours of operation, at 60 Hz repetition rate, 400 ns Pulse
Width. Electric Field: 73 MV/m \cite{adolphsen:ISG8}}
\label{T5RAF-RF}
\end{figure}

\medskip
So far, despite all the unknowns, coarse localization of
breakdowns in structures  have been successful. Acoustic data
obtained with the array of sensors placed onto the structure, like
in Fig.\ref{AET53RAoverall}, are in very good agreement,
Fig.\ref{T53RAFbrkdwn}, with the data obtained by RF analysis,
Fig.\ref{T5RAF-RF}. Coarse localization, by quadrant probing, of
breakdown initiator inside a cell was also successful. A much
finer localization of a breakdown, or a precursor, will require to
have an answer on the propagation (attenuation and scattering) of
phonons in annealed copper.

\section{Conclusion}

Localizing damage, by the mean of ultrasonic waves, is commonly
used in aircraft and building industry. Applying generally this
technique to localize RF breakdowns in accelerators is rather new,
and the NLCTA team has been able to successfully demonstrate its
effectiveness. A more complete demonstration of this contention
will appear in another paper.

\medskip
The next step in understanding breakdown by the use of acoustic
sensor is to have a good knowledge of the propagation of acoustic
waves in annealed copper. This understanding will help on accurate
localization, as is possible in superconductivity cavities
\cite{knobloch}. Being able to do so opens the door to knowledge
about particules contamination and the importance of surface
finish in triggering breakdowns.

\section{Acknowledgments}
I would like to warmly thank all the NLCTA team, which I have
abandon to work on other projects.

\end{document}